\documentclass[%
aip,
amsmath,amssymb,
reprint,%
]{revtex4-2}
\usepackage{amsfonts,amsmath,bm,multirow}
\usepackage[OT4]{fontenc}

\usepackage{graphicx}
\usepackage{xcolor}
\usepackage{epstopdf}
\usepackage{dsfont}
\usepackage{physics}
\usepackage{hyperref}
\usepackage{lipsum}
\usepackage{dcolumn}

\usepackage{natbib}

\usepackage{hf-tikz}

\newcommand{\kp}{\bm{k}\!\vdot\!\bm{p}}
\newcommand{\rr}{\bm{r}}
\newcommand{\kk}{\bm{k}}

\definecolor{bred}{HTML}{e31a1c}
\definecolor{bgreen}{HTML}{33a02c}
\definecolor{bblue}{HTML}{1f78b4}

\definecolor{armygreen}{rgb}{0.29, 0.33, 0.13}

\newcommand{\refa}{$^{\color{blue} a}$}
\newcommand{\refb}{$^{\color{blue} b}$}
\newcommand{\refc}{$^{\color{blue} c}$}
\newcommand{\refd}{$^{\color{blue} d}$}
\newcommand{\refe}{$^{\color{blue} e}$}

\newcommand{\blue}[1]{\textcolor{black}{#1}}

\newcolumntype{L}{D{.}{.}{3,4}}

\makeatletter
\def\@email#1#2{%
	\endgroup
	\patchcmd{\titleblock@produce}
	{\frontmatter@RRAPformat}
	{\frontmatter@RRAPformat{\produce@RRAP{*#1\href{mailto:#2}{#2}}}\frontmatter@RRAPformat}
	{}{}
}%
\makeatother
\begin{document}
	
	\preprint{AIP/123-QED}

	\title {Electronic and spectral properties of Ge$_{1-x}$Sn$_x$ quantum dots}

\author{Krzysztof Gawarecki}
\email{Krzysztof.Gawarecki@pwr.edu.pl}
\affiliation{Institute of Theoretical Physics, Wroc\l aw  University of Science and Technology, Wybrze\.ze Wyspia\'nskiego 27, 50-370 Wrocław, Poland}

\author{Jakub Ziembicki}
\affiliation{Department of Semiconductor Materials Engineering, Wroc\l aw  University of Science and Technology, Wybrze\.ze Wyspia\'nskiego 27, 50-370 Wrocław, Poland}
\author{Paweł Scharoch}
\affiliation{Department of Semiconductor Materials Engineering, Wroc\l aw  University of Science and Technology, Wybrze\.ze Wyspia\'nskiego 27, 50-370 Wrocław, Poland}
\author{Robert Kudrawiec}
\affiliation{Department of Semiconductor Materials Engineering, Wroc\l aw  University of Science and Technology, Wybrze\.ze Wyspia\'nskiego 27, 50-370 Wrocław, Poland}

\date{\today}

	\begin{abstract}
		In this paper, we study theoretically the electron and spectral properties of Ge$_{1-x}$Sn$_x$ systems, including alloys, cubic- and spherical quantum dots. The single-particle electron and hole states are calculated within the $sp^3d^5s^*$ tight-binding approach and used in further modeling of the optical properties. We systematically study the interplay of Sn-driven indirect-direct band-gap transition and the quantum confinement effect in systems of reduced dimensionality. We demonstrate the regime of sizes and compositions, where the ground state in Ge$_{1-x}$Sn$_x$ quantum dot is optically active. Finally, we calculate absorbance spectra in experimentally-relevant colloidal quantum dots and demonstrate a satisfactory agreement with experimental data.

	\end{abstract}

	\maketitle

	\section{Introduction}
	\label{sec:intr}
	
	The nanostructures composed of Ge$_{1-x}$Sn$_x$ alloys became the subject of many experimental and theoretical studies~\cite{Zheng2018}. Such an interest is to a large extent motivated by a composition-tunable character of the Ge$_{1-x}$Sn$_x$ band gap.  In fact, it is known that for the Sn content of a few percent, there is a transition from the indirect to the direct band gap~\cite{Wegscheider1990, Polak2017,Zheng2018, Wirths2015}. In the later regime, the material is attractive for optoelectronics-oriented applications~\cite{Zheng2018}. Furthermore, the high carrier mobility makes it a promising candidate for CMOS fabrication~\cite{Gupta2014,Zhang2022,Liu2023}. 
	
	The concept of Ge$_{1-x}$Sn$_x$ quantum dot (QD) can be realized in several ways. The QDs can be created by a colloidal synthesis for a relatively wide range of the Sn content~\cite{Hafiz2016,Tallapally2018}. Such QDs offer a tunable band gap. They are also low-toxic and they exhibit a narrow size dispersity~\cite{Tallapally2018}. Recently, the new type of Ge$_{1-x}$Sn$_x$ QDs has been reported~\cite{Zhang2022}. For such structures, the optically active QDs appear by diffusion of Sn atoms in GeSn stripes. The QDs synthesized in this way are CMOS compatible~\cite{Zhang2022}. Furthermore, also the gate-defined Ge/GeSn QDs are proposed~\cite{DelVecchio2023}. Such a system is a promising candidate for hole spin qubit, and exhibit a weak hyperfine interaction and strong spin-orbit Rashba coupling~\cite{DelVecchio2023}.
	
     The theoretical approaches utilized in the modeling of Ge$_{1-x}$Sn$_x$ alloys and nanostructures include DFT calculations~\cite{Yin2008,Polak2017,ODonnell2021,Esteves2016}, empirical pseudopotential 
     methods~\cite{Low2012, Gupta2013}, the effective mass approach~\cite{Baira2019,Baira2019b}, multiband $\kp$ methods~\cite{Song2019,Chen2020,pascha}, and the tight-binding model~\cite{Pedersen2010,OHalloran2019}. Since the band-gap regimes of Ge$_{1-x}$Sn$_x$ alloy are strongly affected by strain~\cite{Gupta2013}, in many cases it is crucial to properly represent the material elastic properties in the modeling. This can be achieved within the DFT framework~\cite{Polak2017,OHalloran2019,ODonnell2021}. Furthermore, it was recently shown~\cite{OHalloran2019}, that the \blue{Martin's} Valence Force Field model~\cite{Tanner2019} accurately reproduces the Ge$_{1-x}$Sn$_x$ lattice constant $a(x)$.

     In this work, we present a multi-level modeling for quantum dots based on Ge$_{1-x}$Sn$_x$ alloys. First, we perform DFT simulations (with the MBJLDA functional) to obtain the electronic band structures for unstrained and strained bulk crystals: Ge, $\alpha$-Sn, and GeSn (in the zinc blende structure).  These results constitute a fitting target to find the parameters for the $sp^3d^5s^*$ tight-binding model, including strain exponents. We use this model to find the single-particle electron and hole states for cubic Ge$_{1-x}$Sn$_x$ boxes and spherical quantum dots. We investigate the interplay of the indirect to direct band-gap transition and the quantum confinement effect. We emphasize that for small systems, much more Sn is required to obtain the direct band-gap regime, compared to the alloy case. For the spherical QDs, we calculate absorption spectra and compare them to experimental data for colloidal QD systems. In this kind of simulation, we include the effect of Coulomb interaction via the diagonal shift in the energies (the Hartree approximation). We demonstrate a satisfactory agreement between the theoretical and experimental results.
	
     The paper is organized as follows. In Sec.~\ref{sec:model}, we define the computational model. The tight-binding parameters and comparison between the band structures obtained via the tight-binding and the DFT are given in Sec.~\ref{sec:bulk}. The fundamental concepts related to Ge$_{1-x}$Sn$_x$ systems in the presence of confinement are discussed in subsequent Sec.~\ref{sec:box}. In Sec.~\ref{sec:colloidal}, we present the results for the spherical QDs. Finally, Sec.~\ref{sec:conclusions} contains the conclusions.

	\section{Model}
	\label{sec:model}
	\subsection{Tight binding model}
	
	To model the electron and hole states, we use the sp$^3$d$^5$s$^*$ tight-binding model within the approximation of the nearest neighbors (NN). The Hamiltonian~\cite{Slater1954,Jancu1998,zielinski12} can be written as
	\begin{align*}
	H = & \sum_{i} \sum_{\alpha}  E^{(i)}_{\alpha} a^{\dagger}_{i\alpha} a_{i\alpha}  + \sum_{i} \sum_{j \neq i}^{\mathrm{NN}(i)} \sum_{\alpha,\beta} t^{(ij)}_{\alpha \beta} \, a^{\dagger}_{i\alpha} a_{j\beta} \\ & +  \sum_{i} \sum_{\alpha,\beta} \Delta^{(i)}_{\alpha\beta} \, a^{\dagger}_{i\alpha} a_{i\beta} ,
	\end{align*}
	where $a^{\dagger}_{i\alpha}$/$a_{i\alpha}$ is the creation/anihilation operator of the $\alpha$-th orbital at the atomic site of the position $\bm{R}_i$. The $E^{(i)}_{\alpha}$ are on-site energies, $t^{(ij)}_{\alpha \beta}$ is the hopping parameter for a given pair of atoms and orbitals, and $\Delta^{(i)}_{\alpha\beta}$ refers to the spin-orbit coupling~\cite{Chadi1977}. 	
	For the calculations in nanostructures, the eigenstates of the Hamiltonian can be written in form
	\begin{equation}
		\ket{\Psi_{n}} = \sum_{i,\alpha} \varphi_{n,\alpha}(\bm{R}_i) \ket{\bm{R}_i; \alpha},
	\end{equation}
	where $\varphi_{n,\alpha}(\bm{R}_i)$ are complex coefficients and $\ket{\bm{R}_i; \alpha}$ is the $\alpha$-th atomic orbital localized at the $i$-th atom.
	\begin{table}
		\caption{\label{tab:params_strain} The material parameters: the lattice constant $a$ (in \AA), the force constants $\alpha$ and $\beta$ (in $10^{12}$~dyne/cm), and the bulk elastic constants (in $10^3$~dyne/cm$^2$). The elastic constants for Ge and Sn are taken from Ref.~\cite{Polak2017} and for GeSn from Ref.~\cite{WittekBSC}.}
		\begin{ruledtabular}
			\begin{tabular}{lllccc}
				& Ge & Sn & GeSn & source\\
				\hline \\[-0.5em]
				$a$ & 5.647 & 6.479 & 6.071 & DFT \\[1.1pt]
				$\alpha$  & 37.13 & 27.54 & 35.65 & calc.\\[1.1pt]
				$\beta$  & 10.59 & 5.51 &  7.60 & calc. \\[1.1pt]
				\hline\\[-0.5em]
				$C_{11}$  & 12.2 & 6.8 & 9.63 & DFT \\[1.1pt]
				$C_{12}$  & 4.7 & 3.4 & 4.62 & DFT \\[1.1pt]
				$C_{44}$  & 8.6 & 5.3 & 4.55 & DFT \\[1.1pt]
				\hline\\[-0.5em]
				$C_{44}$  & 5.83 & 2.83 & 4.13 & calc. \\[1.1pt]
				
			\end{tabular}
		\end{ruledtabular}
	\end{table}

        \subsection{Nanostructures}
        \label{subsec:nanostructures}
	\begin{figure}[tb]
		\begin{center}
			\includegraphics[width=.48\textwidth]{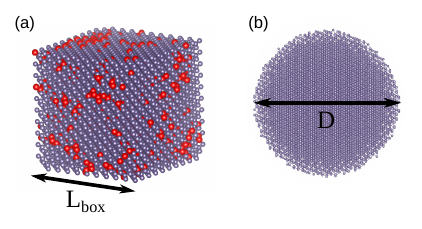}
		\end{center}
		\caption{\label{fig:structures} (a-b) Schematic pictures of the considered Ge$_{1-x}$Sn$_x$ structures. The violet and red balls corresponds to Ge and Sn atoms, respectively.}
	\end{figure}
        We performed the nanostructure modeling with the tight-binding model. Such an approach is widely used in the modeling of colloidal quantum dots~\cite{Jain2016}, and was successful in the description of Ge nanocrystals~\cite{Lee2009}. In our work, we do the calculations with an atomistic grid containing up to several millions of atoms.
        For the Ge$_{1-x}$Sn$_x$ systems, the atomic species are generated randomly with $x$ referring to a probability of finding Sn atom at a given atomic site. We consider two geometries of the system: the cubic box [Fig.~\ref{fig:structures}(a)], and spherical QDs [Fig.~\ref{fig:structures}(b)].
        
        All simulations start with the cubic domain with periodic boundary conditions. The atomic positions are (uniformly) scaled to obtain the effective alloy lattice constant of $\widetilde{a}(x) = (1 - x) a_{\mathrm{Ge}} + x a_{\mathrm{Sn}} - b x (1 - x)$, where we took $b = -0.083$~\AA~\cite{Polak2017}.  To minimize local strain resulting from inhomogeneous Sn concentration, we perform further optimization of the atomic positions (see Sec.~\ref{subsec:minstrain}), while the overall size of the domain is fixed. 
 
	For the calculation of states, we use two types of boundary conditions:
	\begin{enumerate}
		 \item[BC1] The periodic boundary conditions in all directions. In this case, we keep the initial cubic domain. No dangling bonds are present in the calculations.
   
		 \item[BC2] No periodic boundary conditions in any direction. After the local strain minimization, a smaller box or sphere is ``carved out" from the central part of the initial domain. In order to prevent the appearance of surface states with energies inside the band gap, dangling bonds at the surface are passivated as described in Ref.~\cite{Lee2004}, where we took the parameter $\delta = 15$~eV.
		 
	 \end{enumerate}

	\subsection{Local minimization of strain}
        \label{subsec:minstrain}
	To model bulk material systems in the presence of hydrostatic and biaxial strain, the atomic positions in the unit cell are altered~\cite{Kleinman1962,Gawarecki2019b}. In the case of Ge$_{1-x}$Sn$_x$ nanostructures, strain is inherently present due to lattice mismatch between Ge and Sn crystals. To obtain resulting atomic positions, we perform calculations in the standard \blue{Keating's} valence force field (VFF) model with the energy functional~\cite{Pryor1998b}
	\begin{align*}
		U &= \frac{3}{16} \sum_{i} \sum_{j}^{\mathrm{NN}(i)} \frac{ \alpha_{ij} }{d_{ij}^{2}} \left( \bm{r}_{ij}^{2} - d_{ij}^{2} \right)^{2} \\
		& \phantom{=} + \frac{3}{16} \sum_{i}  \sum_{j,k>j}^{\mathrm{NN}(i)}  \frac{ \beta_{ij} + \beta_{ik} }{d_{ij} d_{ik}}  \left( \bm{r}_{ij} \bm{r}_{ik} - d_{ij} d_{ik} \cos{\theta_{0}} \right)^{2},	
	\end{align*}
	where $\bm{r}_{ij} = {\bm{R}_j - \bm{R}_i}$ and $d_{ij}$ are the actual and the unstrained interatomic distances between the atoms at the $i$-th and $j$-th node; the ideal bond angle is given by $\cos{\theta_0} = -1/3$.
	The $U$ functional minimization was performed using parallel PETSC TAO library~\cite{petsc-web-page}.
	The values of parameters $\alpha$ and $\beta$ are adjusted to reproduce $C_{11}$ and $C_{12}$ bulk elastic constants with the formulas $\alpha = a (C_{11} + 3 C_{12} )/4$ and $\beta = a(C_{11}-C_{12})/4$ ~\cite{Keating1966}, where $a$ is the lattice constant. One should note, that the accuracy of $C_{11}$ and $C_{12}$ is preferred over the $C_{44}$, which is not a target for the fitting~\cite{Zielinski2012}. This approach is justified by the fact, that for the considered structures the hydrostatic strain is much more important than the shear strain. 
	The value of $C_{44}$ obtained by the formula~\cite{Pryor1998b}
	\begin{equation}
		C_{44} = \frac{4 \alpha \beta}{a(\alpha + \beta)}
	\end{equation}
	is underestimated compared to the DFT data (see Table~\ref{tab:params_strain}). 

         As described in Sec.~\ref{subsec:nanostructures}, the initial atomic positions are forced to reproduce the alloy lattice constant $a(x)$, which is fitted to the DFT data~\cite{Polak2017}. The VFF simulations are performed to tune local atomic positions, while the overall size of the initial domain is kept fixed. The motivation for that procedure is the fact, that the standard VFF model underestimates the alloy lattice constant for Ge$_{1-x}$Sn$_x$. This can be related to the relative simplicity of the VFF model (only two parameters) and the large lattice mismatch between Ge and Sn. A higher accuracy can be obtained within the \blue{Martin's} VFF~\cite{Tanner2019} containing more parameters, which implementation is out of the scope of our paper.
	
	We include the strain into the tight-binding model by changing the bond angles, adding a shift to the on-site energy in the $d$-shell (with the $b_\mathrm{d}$ parameter), and rescaling the two-center parameters $t^{(ij)}_{\alpha \beta} \rightarrow t^{(ij)}_{\alpha \beta} (d_{ij}/\abs{\bm{r}_{ij}})^{n^{(ij)}_{\alpha\beta\kappa}}$ using the strain exponents~\cite{Ren1982,Jancu1998,Jancu2007,Gawarecki2019b}. \blue{The $b_\mathrm{d}$ parameter improves the treatment of [001]-like uniaxial strain~\cite{Jancu2007}. We note that more advanced schemes were proposed~\cite{Niquet2009, Jancu2007}, which offer enhanced accuracy (in particular for shear strain).} 

 	\subsection{Absorbance spectrum}
	\label{sec:abs}
	 To model the absorption spectra in colloidal QDs, we calculate the oscillator strengths between the discrete single-particle states
	 \begin{equation}
	 	f_{mn} = -\frac{2}{3 m_0} \sum_{i=x,y,z} \frac{ \abs{(P_i)_{mn}}^2 }{ E_m - E_n},
	 \end{equation}
	 where $E_k$ are the energies of states from the valence band (the $m$ index) or the conduction band (indexed by $n$). 
	 The momentum matrix elements $(P_i)_{mn}$ between the states are calculated using the Hellmann-Feynman theorem\cite{Feynman1939, LewYanVoon1993, eissfeller12}, as widely described in Ref.~\cite{Gawarecki2020}. \blue{To have states with a well-defined axial projection of spin}, we perform the simulations for subsequent Fig.~\ref{fig:colloidal_abs} in a small magnetic field $B_z = 0.1$~T. 
      Neglecting the frequency dependence for the refractive index $n$, the absorbance is proportional to~\cite{Yu2010}
	 \begin{equation}
	 	\label{eq:absPropTo}
	 	\alpha(\omega) \propto {\omega \Im{\epsilon(\omega)}},
	 \end{equation}	 
	 with the imaginary part of the dielectric function 
	 \begin{equation}
	 	\Im{\epsilon(\omega)} \propto \sum_{m,n} \frac{f_{mn} \, \omega \, \gamma}{(\omega^2_{mn} - \omega^2)^2 + \omega^2 \gamma^2},
	 \end{equation}
	 where $\omega_{mn} = (E_n - E_m - V^\mathrm{eh}_{nmmn})/\hbar$ and $\gamma$ is the decrement parameter responsible for line broadening. As the simulations for the absorbance are performed with a very large number of states, the standard configuration-interaction scheme would be very demanding computationally. Instead, we take into account the impact of the Coulomb coupling via the diagonal matrix elements between the electron and hole states~\cite{Schulz2006} \blue{
	 \begin{align*}
	 	V^\mathrm{eh}_{nmmn} & \approx \frac{{e}^{2}}{4 \pi \epsilon_{0} \epsilon_{r}} \sum_{i,j \neq i}^{N_a} \sum_{\alpha,\beta}^{20} \\ & \times   \frac{\varphi^{(e)*}_{n,\alpha}(\bm{R}_{i}) \varphi^{(h)*}_{m,\beta}(\bm{R}_{j}) \varphi^{(h)}_{m,\beta}(\bm{R}_{j}) \varphi^{(e)}_{n,\alpha}(\bm{R}_{i})}{\abs{\bm{R}_{i} - \bm{R}_{j}}},
 	 \end{align*}
	  where $\varphi^{(e/h)}_{n,\alpha}$ are the single-particle electron/hole wavefunctions, $\epsilon_{r} = -7.64701 x^2 + 16.5545x + 18.18941$ is the dielectric constant of Ge$_{1-x}$Sn$_x$~\cite{Chen2023}.}
	 A detailed description of the calculation of such matrix elements is given in the Appendix of Ref.~\cite{Gawarecki2023a}. One should note, that moving from the picture of valence band states to the picture of hole states requires the time reversal operation.

	\section{Bulk materials: parameter fitting}
	\label{sec:bulk}
	Calculations in the framework of the tight-binding model require a comprehensive parameter set including on-site energies, hopping integrals, and strain exponents~\cite{Slater1954,Jancu1998,Ren1982}. In the case of Ge, the present literature offers several parameter sets for the $sp^3d^5s^*$ TB model~\cite{Jancu1998, Boykin2004, Niquet2009}. On the other hand, the sources for Sn are scarcer, although the parameters are given for the $sp^3$ model~\cite{Pedersen2010}. One should note that in the case of Ge$_{1-x}$Sn$_x$ nanostructure, the hopping integrals between Ge and Sn atoms are needed~\cite{OHalloran2019}. To find parameters for the $sp^3d^5s^*$ TB model that are consistent for all the considered materials, we performed multi-level fitting to a coherent set of DFT data. 
	\begin{figure*}[tb]
		\begin{center}
			\includegraphics[width=.9\textwidth]{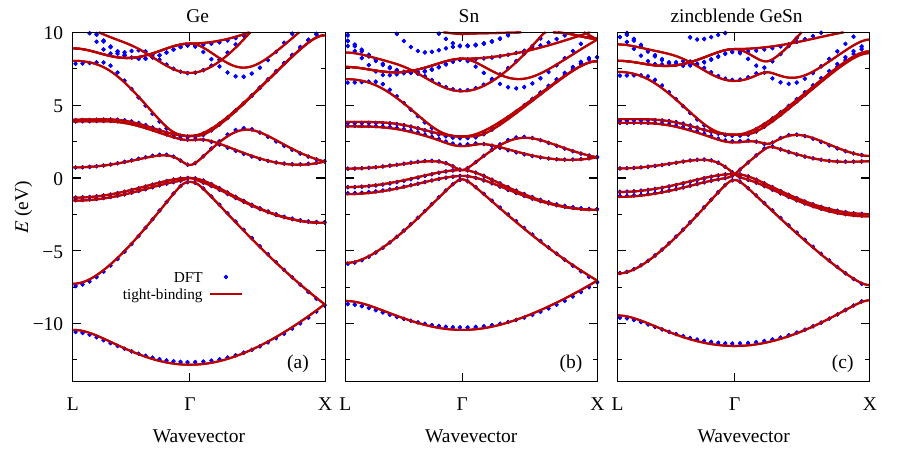}
		\end{center}
		\caption{\label{fig:bulk_bs} Band structures for Ge, Sn, and GeSn calculated within the DFT and the TB approaches.}
	\end{figure*}

	We calculated the band structures for unstrained Ge, Sn and GeSn crystals within the DFT approach using VASP code and PAW method~\cite{VASP,VASP2,VASP3,VASP_PAW}. Geometrical optimization was performed with LDA functional and electronic structure calculation with the MBJ functional~\cite{tran_blaha_2009}. The artificial GeSn crystal (in the zincblende atomic structure)~\cite{OHalloran2019} was calculated to further extract the hopping integrals between the Ge and Sn atoms. The target band structures calculated via DFT are shifted by the valence band offset (VBO), which is taken VBO(Ge) = $0.0$~eV and VBO(Sn) = $0.55$~eV~\cite{Polak2017}. For GeSn crystal, we take the averaged value VBO(GeSn) = $0.275$~eV. 
	
	To find the TB parameters, we performed a fitting using LMFIT~\cite{Newville2014} Python library within the least squares method. We started with parameters for Ge from Ref.~\cite{Jancu1998} modified by some random noise. To avoid finding local minima, the procedure was performed for multiple realizations of random noise. 
	In the fitting, we assigned the greatest weights to the target points near the Fermi level and in proximity of the high-symmetry points ($\Gamma$, $L$, and $X$). In the case of GeSn, we take the same on-site parameters as in the Ge and Sn crystals (similarly to Ref.~\cite{Niquet2009}, where such an approach was used to find GeSi parameters). 
	The obtained parameter set is given in Table~\ref{tab:TB_bulk} and the resulting band structures for unstrained crystals are shown in Fig.~\ref{fig:bulk_bs}. As one can see, there is a very good agreement between the TB and DFT results for the bands of interest. The parameters listed in Table~\ref{tab:TB_bulk} reproduce also the valence band offsets, as described above. To find the strain exponents $n^{(ij)}_{\alpha\beta\kappa}$ and the $b_\mathrm{d}$ parameters, we performed additional DFT calculations in the presence of hydrostatic and axial strain.  Here, the fitting was performed only for high symmetry points with the greatest weights for the band edges close to the Fermi level. 
	Also in this case, we obtain a satisfactory agreement between the models. \blue{In the case of $b_\mathrm{d}$ for Sn, the fitting gave very small value, which we approximated by 0.0.}
	
	\blue{The comparison between our band structure parameters and the values reported in literature is shown in Table~\ref{tab:params_comparison}. We extracted the Kane energy ($E_P$) and the reduced Luttinger parameters $\gamma'_{1-3}$ by fitting the TB results to the eight-band $\kp$ model~\cite{Winkler2003} in the vicinity of the $\Gamma$ point. Then, the Luttinger parameters are obtained via the perturbative formulas~\cite{Winkler2003}
	\begin{align*}
		\gamma_1 &= \gamma'_1 + \frac{E_P}{3 E^\mathrm{(\Gamma)}_\mathrm{g}}, \\
		\gamma_2 &= \gamma'_2 + \frac{E_P}{6 E^\mathrm{(\Gamma)}_\mathrm{g}}, \\
		\gamma_3 &= \gamma'_3 + \frac{E_P}{6 E^\mathrm{(\Gamma)}_\mathrm{g}}. 
	\end{align*}
	As one can see, the obtained Kane energies are lower than the reported values, which means that our band structures are more flat. This behavior is a known shortcoming of the MBJLDA approach~\cite{Kim2010}.}

	\begin{table}
		\caption{\label{tab:TB_bulk} The tight binding parameters (on-site energies, the spin-orbit coupling parameter, hopping integrals and strain-related exponents) for Ge, Sn, and GeSn crystals. All the parameters, except dimensionless exponents ($n_{ij\kappa}$), are given in eV. For GeSn, the two-center parameters $A B \xi$ (where $\xi$ = $\{\sigma,\pi,\delta\}$) refer to the bracket notation $(V_1,V_2)$ in the following way: $A_\mathrm{Ge} B_\mathrm{Sn} \xi = V_1$ and $A_\mathrm{Sn} B_\mathrm{Ge} \xi = V_2$; e.g. $s_\mathrm{Ge}p_\mathrm{Sn}\sigma = 2.4987$~eV and $s_\mathrm{Sn}p_\mathrm{Ge}\sigma = 2.3729$~eV.
		}
		\begin{ruledtabular}
			\begin{tabular}{lllccc}
				 & Ge & Sn & GeSn\\
				\hline \\[-0.8em]
				${E_{\mathrm{s}}}$   & -4.5429 & -3.7460 &  \\[1.1pt]
				${E_{\mathrm{p}}}$   & 3.3584  &  2.9219 &  \\[1.1pt]
				${E_{\mathrm{d}}}$   & 12.5970 & 10.1730 &  \\[1.1pt]
				${E_{\mathrm{s}^*}}$ & 18.4900 & 16.0590 &  \\[1.1pt]
				$\Delta$             & 0.3248  &  0.7539 &  \\[1.1pt]
				\hline \\[-0.8em]
				$ss\sigma$ \hfill      &  -1.6246 & -1.2139 & -1.3625 \\[1.1pt]
				$s^*s^*\sigma$ \hfill  &  -3.2654 & -2.9841 & -3.1255 \\[1.1pt]
				$ss^*\sigma$ \hfill    &  -1.4290 & -1.2928 & ( -1.4834, -1.3001)\\[1.1pt]
				$sp\sigma$ \hfill      &   2.6587 &  2.1576 & (  2.4987,  2.3729)\\[1.1pt]
				$s^*p\sigma$ \hfill    &   2.0262 &  1.7421 & (  2.6935,  1.4506)\\[1.1pt]
				$sd\sigma$ \hfill      &  -1.5562 & -1.2690 & ( -1.3930, -0.8382)\\[1.1pt]
				$s^*d\sigma$ \hfill    &   1.5649 &  1.0671 & (  1.7832,  0.9477)\\[1.1pt]
				\hline \\[-0.8em]
				$pp\sigma$ \hfill      &   3.5386 &  2.8134 &  3.1250 \\[1.1pt]
				$pp\pi$ \hfill         &  -1.2945 & -1.0136 & -1.1298 \\[1.1pt]
				$pd\sigma$ \hfill      &  -1.1915 & -0.8312 & ( -0.8584, -1.1712)\\[1.1pt]
				$pd\pi$ \hfill         &   2.0446 &  1.5726 & (  2.1277,  1.2724)\\[1.1pt]
				\hline \\[-0.8em]
				$dd\sigma$ \hfill      &  -1.2509 & -0.7432 & -0.9079 \\[1.1pt]
				$dd\pi$ \hfill         &   1.9585 &  1.2726 &  1.4888\\[1.1pt]
				$dd\delta$ \hfill      &  -1.3953 & -1.0603 & -1.3145\\[1.1pt]
				 \hline \\[-0.8em]
				$n_{ss\sigma}$ \hfill     &   4.2013&  4.0870 &  3.6640 \\[1.1pt]
				$n_{s^*s^*\sigma}$ \hfill &   0.0000&  0.0000 &  0.0000 \\[1.1pt]
				$n_{ss^*\sigma}$ \hfill   &   0.0000&  0.0000 &  0.0000 \\[1.1pt]
				$n_{sp\sigma}$ \hfill     &   3.3615&  4.2412 &  2.5245 \\[1.1pt]
				$n_{s^*p\sigma}$ \hfill   &   0.0000&  6.9997 &  0.4633 \\[1.1pt]
				$n_{sd\sigma}$ \hfill     &   6.8670&  4.4558 &  4.8123 \\[1.1pt]
				$n_{s^*d\sigma}$ \hfill   &   2.0000&  2.0000 &  2.0000 \\[1.1pt]
				$n_{pp\sigma}$ \hfill     &   1.5985&  1.9464 &  2.7408 \\[1.1pt]
				$n_{pp\pi}$    \hfill     &   2.6281&  2.8530 &  2.6646 \\[1.1pt]
				$n_{pd\sigma}$ \hfill     &   2.7964&  1.8281 &  0.0000 \\[1.1pt]
				$n_{pd\pi}$    \hfill     &   0.0000&  2.3247 &  0.0000 \\[1.1pt]
				$n_{dd}$ \hfill           &   4.2082&  0.0000 &  4.6073 \\[1.1pt]

				\hline \\[-0.8em]
				$b_{d}$ \hfill & 0.0929 & 0.0000 &
			\end{tabular}
		\end{ruledtabular}
	\end{table}
	\begin{table}
	\caption{\label{tab:params_comparison} The comparison of band structure parameters. The energy splittings are calculated from $E^\mathrm{(L)}_\mathrm{g} = E(L_\mathrm{6c}) - E(\Gamma_\mathrm{8v})$, $E^\mathrm{(\Gamma)}_\mathrm{g} = E(\Gamma_\mathrm{7c}) - E(\Gamma_\mathrm{8v})$, $\Delta = E(\Gamma_\mathrm{8v}) - E(\Gamma_\mathrm{7v})$.}
	\begin{ruledtabular}
	\begin{tabular}{lllccc}

		& \multicolumn{2}{c}{Ge} & \multicolumn{2}{c}{Sn}\\[1.1pt]
		& This work & Other works & This work & Other works \\[1.1pt]
		\hline \\[-0.5em]
		$E^\mathrm{(L)}_\mathrm{g}$ & 0.725 & 0.744\refd, 0.76\refa & 0.098 & 0.14\refa  \\[1.1pt]
		$E^\mathrm{(\Gamma)}_\mathrm{g}$ & 0.890 & 0.898\refd, 0.90\refa & -0.396  & -0.42\refa  \\[1.1pt]
		$\Delta$ & 0.2683 & 0.29\refa & 0.636 & 0.8\refa \\[1.1pt]
		\hline\\[-0.5em]
		$E_{\mathrm{P}}$ & 20.52 & 26.3\refb & 16.23 & 23.8\refb, 18.8\refe \\[1.1pt]
		$\gamma_{1}$  & 10.55 & 13.35\refb, 13.38\refc & -10.56 & -14.97\refb, -12.0\refe \\[1.1pt]
		$\gamma_{2}$  & 3.21 & 4.25\refb, 4.24\refc & -7.36 & -10.61\refb, -8.45\refe \\[1.1pt]
		$\gamma_{3}$  & 4.55 & 5.69\refb, 5.69\refc & -6.13 & -8.52\refb, -6.84\refe \\[1.1pt]
		\toprule\\[0.5pt]
		\multicolumn{5}{p{0.97\linewidth}}{\rule{0pt}{1.em} \refa Values from Ref.~\onlinecite{Madelung2004}.} \\
		\multicolumn{5}{p{0.97\linewidth}}{\rule{0pt}{1.em} \refb Values from Ref.~\onlinecite{Lawaetz1971}.} \\
		\multicolumn{5}{p{0.97\linewidth}}{\rule{0pt}{1.em} \refc Values from Ref.~\onlinecite{Hensel1974}.} \\
		\multicolumn{5}{p{0.97\linewidth}}{\rule{0pt}{1.em} \refd Values from Ref.~\onlinecite{Zwerdling1959}.} \\
		\multicolumn{5}{p{0.97\linewidth}}{\rule{0pt}{1.em} \refe Values from Ref.~\onlinecite{Brudevoll1993}.} 		
	\end{tabular}
	\end{ruledtabular}
	\end{table}

	\section{Single-particle states in a cubic box}
	\label{sec:box}
	\begin{figure}[tb]
		\begin{center}
			\includegraphics[width=.48\textwidth]{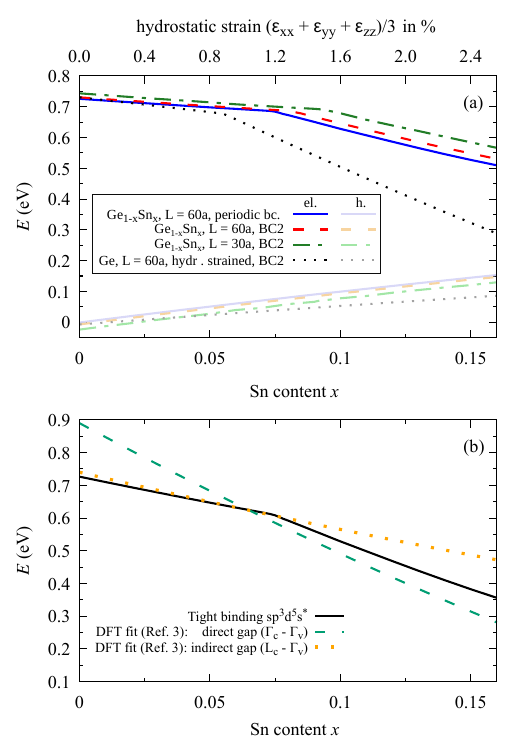}
		\end{center}
		\caption{\label{fig:box} (a) Energy levels corresponding to the lowest conduction-band states and the highest valence-band states for Ge$_{1-x}$Sn$_{x}$ boxes as a function of composition (the bottom scale); and the results for Ge box in the presence of tensile strain (the scale on the top) reproducing Ge$_{1-x}$Sn$_{x}$ lattice constant. (b) The energy gap calculated from the energy difference of the lowest cb. and the highest vb. single-particle states in the Ge$_{1-x}$Sn$_{x}$ box with periodic boundary conditions, and the lines fitted to the DFT results (Ref.~\cite{Polak2017}).}
	\end{figure}
    We calculated the lowest energy levels for electron and hole (in fact, the valence band electron) confined in a Ge$_{1-x}$Sn$_x$ cubic box. For the sake of comparison, we consider also a pure Ge box in the presence of external tensile, hydrostatic strain. The results are presented in Fig.~\ref{fig:box}(a). We investigate cubic structures 
    for the edge lengths $L_{\mathrm{box}} = 60a$ and $L_{\mathrm{box}} = 30a$; and two types of boundary conditions.
    
     Before moving to the analysis of the numerical results, to illustrate the basic concepts, let us first consider the simplest case of an infinite ($L_\mathrm{box} \rightarrow \infty$), unstrained Ge box. One should recall that Ge bulk material has the indirect band gap ($L_{6c}$ -- $\Gamma_{8v}$), where the lowest conduction band (cb) has a minimum energy at the $L$ point [see Fig.~\ref{fig:bulk_bs}(a)]. In consequence, the electron ground state is an arbitrary superposition of the Bloch functions from the eight equivalent $L$ points $\mathcal{L} = \qty{ \pi / a (\pm 1, \pm 1, \pm 1)}$  
    \begin{equation}
   	   	\label{eq:psi0exp_ideal}
    	\Psi^{(\infty)}_0(\rr) = \sum_{\bm{k} \in \mathcal{L}} c_{\bm{k}} \, e^{i \bm{k} \bm{r}} \, u_{\mathrm{cb},\kk}(\rr),
    \end{equation}
	where $c_{\bm{k}}$ are complex coefficients and $u_{\mathrm{cb},\kk}$ is the periodic part of the Bloch wave function. The energy of $\Psi^{(\infty)}_0$ is the bulk value, which in the TB is $E_\mathrm{cb}^{(L)} \approx 0.725$~eV. 
	
	We performed calculations for $L_{\mathrm{box}} = 60a$ and the periodic boundary conditions (BC1). This can be interpreted as a supercell bulk simulation~\cite{OHalloran2019} (in a supercell containing $60\times60\times60$  cubic cells) taken at $\bm{k} = 0$. In such a model, the band folding takes place. For a pure Ge ($x=0$) box and no external strain, we get the ground state energy with the value of $E_\mathrm{cb}^{(L)}$ (the difference below $1 \mu$eV, which was set as a tolerance of the numerical diagonalization). 
	
	In a finite-size system, the energy increases due to the confinement. In fact, for Ge boxes of $L_{\mathrm{box}} = 30a$ and $L_{\mathrm{box}} = 60a$ calculated with BC2, the ground state energies are $0.743$~eV and $0.730$~eV, respectively. The difference is relatively small since the cb band is flat (has a large effective mass) near the $L$ points in the BZ [see Fig.~\ref{fig:bulk_bs}(a)]. 
	For a finite-size box of a crystal material, 
	the ground state can be expanded into the bulk wave functions via
    \begin{equation}
   	\label{eq:psi0exp}
		\Psi_0(\rr) =  \sum_{n} \int_{\mathrm{BZ}} C_n(\bm{k}) \, e^{i \bm{k} \bm{r}} \, u_{n,\kk}(\rr) \dd{ \kk},
	\end{equation}	
	where $n$ is the band index, $C_n(\bm{k})$ are distribution functions, and integration covers the full BZ. One can expect, that for reasonably large structures the dominant contribution in Eq.~\ref{eq:psi0exp} comes from $n=\mathrm{``cb"}$ and $\bm{k}$'s near the $L$ points, tending to reproduce Eq.~\ref{eq:psi0exp_ideal}. 
	
	\begin{figure*}[tb]
		\begin{center}
				\includegraphics[width=.9\textwidth]{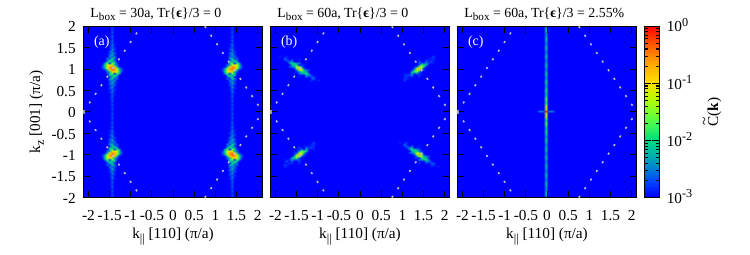}
			\end{center}
		\caption{\label{fig:contrib} (a-c) The total contribution $\widetilde{C}(\bm{k})$ from a given part of the BZ to the ground state in the pure Ge box. The ($1\bar{1}0$) plane is presented and the dotted line marks the edge of the first BZ. In the case of the strained box (c), the lattice constant $a$ is rescaled.}
	\end{figure*}
	 To further examine the states confined in the boxes, we calculated the coefficients
	\begin{equation}
		C_n(\bm{k}) = \sum_{j} \sum_{\alpha} \varphi_{0,\alpha}(\bm{R}_j) e^{-i \bm{k} \bm{R}_j} \, \braket{u_{n\kk}}{\bm{R}_j;\alpha},
	\end{equation}
	where $\ket{u_{n\kk}}$ are expressed via combinations of the atomic orbitals at each type of the lattice site, resulting from the diagonalization of the bulk TB Hamiltonian at $\kk$. To estimate the total contribution coming from a given part of the BZ, we calculate 
	\begin{equation*}
	\widetilde{C}(\bm{k}) = \sqrt{ \sum_n^{40} \abs{C_n(\bm{k})}^2},
	\end{equation*}
	where $40$ is the number of bands accounted for within the sp$^3$d$^5$s$^*$ tight-binding model. The results of $\widetilde{C}(\bm{k})$ for the ($1\bar{1}0$) plane in the Brillouin Zone are shown in Figs.~\ref{fig:contrib}(a-c). We considered pure Ge boxes for the boundary conditions BC2. Fig.~\ref{fig:contrib}(a) and Fig.~\ref{fig:contrib}(b) show the results for unstrained boxes of $L_{\mathrm{box}} = 30a$ and $L_{\mathrm{box}} = 60a$, respectively. In both cases, the distributions resemble the idealistic case of Eq.~\ref{eq:psi0exp_ideal}, yet the smaller structure exhibits a broader distribution in the $\kk$-space, as one can expect.
	
	From the point of view of virtual crystal approximation (VCA), in Ge$_{1-x}$Sn$_x$ alloy the \blue{$\Gamma_{7c}$} decreases with $x$ more rapidly than $L_{6c}$~\cite{pascha,Yin2008}. In consequence, the band gap becomes direct (\blue{$\Gamma_{7c}$} -- $\Gamma_{8v}$) for Sn concentration exceeding some critical value $x_\mathrm{c}$. For the unstrained material, this critical composition can be estimated to be $5$--$11$~\%~\cite{Yin2008,Polak2017,pascha}. 
	
	We performed simulations for the Ge$_{1-x}$Sn$_x$ composition dependence of the electron ground state energy in the box [Fig.~\ref{fig:box}(a)]. The change in the line slope indicates the transition between the indirect- to direct-band-gap regime. In the case of the periodic boundary conditions BC1 and $L_{\mathrm{box}} = 60a$, we obtain $x_c \approx 7.5$\%, which is close to the alloy value of $x_\mathrm{c} \approx 6.5$~\% found within the DFT approach~\cite{Polak2017} (see Fig.~\ref{fig:box}(b) for a direct comparison).
	On the other hand, for the BC2 the transition takes place for a considerably larger Sn content than for the alloy. This difference is mainly due to the fact that the effective mass at $\Gamma_{7c}$ is much smaller than the one at the $L$ point. Thus in the presence of the confinement, the electron states composed of the Bloch functions from the vicinity of $\Gamma_{7c}$ are shifted up in the energy. \blue{Such a behavior was invoked in Ref.~\onlinecite{Niquet2000} to explain the size-dependence for the radiative lifetime in Ge nanocrystals.}
     The transition between the band-gap regimes is gradual, and some hybridization occurs near the $x_c$, which is consistent with Refs.~\cite{Eales2019,OHalloran2019}.
	Finally, we calculated the valence band states. In this case, no transitions appear and the energies are increasing along with the Sn content. The slope of this dependence is governed by the effect of strain and the VBO between Ge and Sn.
	
	The pure Ge material changes its band gap character under tensile strain~\cite{Niquet2009, Liu2014b, Sukhdeo2016}. Also, the direct-indirect band gap transition in Ge$_{1-x}$Sn$_x$ to a large extent relies on increasing interatomic distances between Ge atoms. Such a spatial expansion is governed by Sn, which has a larger lattice constant than Ge (6.479~\AA \, vs. 5.647~\AA). For the sake of comparison, we calculated energies in Ge box, which was artificially expanded to match the Ge$_{1-x}$Sn$_x$ structure. This corresponds to a tensile hydrostatic strain
	\begin{equation*}
		\epsilon_{xx} = \epsilon_{yy} = \epsilon_{zz} = \frac{\widetilde{a}(x) - a_\mathrm{Ge}}{a_\mathrm{Ge}}.
	\end{equation*} 
	The boundary conditions BC2 were imposed. The results are shown in Fig.~\ref{fig:box}(a) with the corresponding scale on the top. As for the compositional dependence, there is the indirect to direct band gap transition~\cite{Niquet2009,Liu2014b, Sukhdeo2016}. While for the unstrained Ge box, the electron ground state is mainly built by the $L$-point-like contributions [Fig.~\ref{fig:contrib}(b)], for tensile hydrostatic strain above about $\epsilon_{xx} = \epsilon_{yy} = \epsilon_{zz} = 0.89$\% the state has the $\Gamma$-like character. This is clearly seen in Fig.~\ref{fig:contrib}(c), where the box is strained by 2.55\%. In this case, one can also observe the contributions coming from the $\Delta$ valley ($\Gamma$ -- $X$). In contrast to the  Ge$_{1-x}$Sn$_x$ case, the transition between the regimes is sharp. This can be related to the fact, that the Ge box strained hydrostatically has the $\mathrm{O}_\mathrm{h}$ symmetry, while for the Ge$_{1-x}$Sn$_x$ systems the symmetry is completely lifted (to $C_1$) by the random appearance of atoms.

	\section{Spherical quantum dots}
	\label{sec:colloidal}
	\begin{figure}[tb!!]
		\begin{center}
			\includegraphics[width=.48\textwidth]{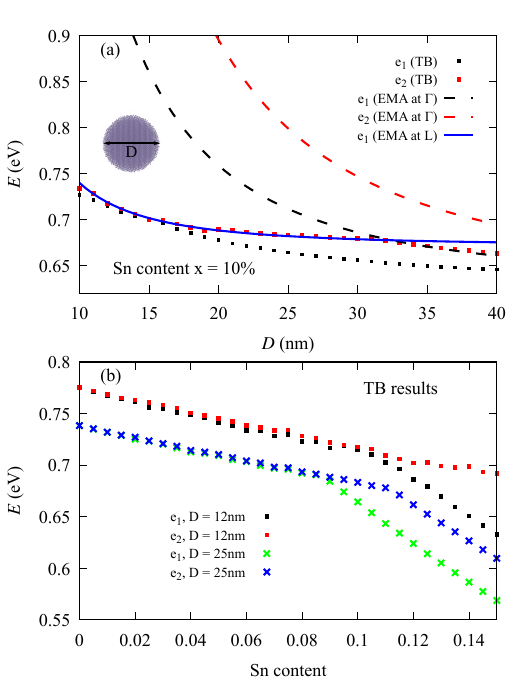}
		\end{center}
		\caption{\label{fig:colloidal_en} (a) Energy of the lowest electron states in spherical QDs, calculated using the TB and the effective mass methods, as a function of the dot diameter ($D$); (b) the dependence of the lowest energy levels on the QD composition for the two fixed dot diameters. }
	\end{figure}

	\begin{figure}[tb!!]
		\begin{center}
			\includegraphics[width=.48\textwidth]{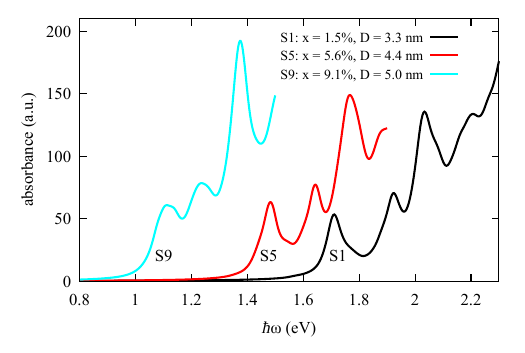}
		\end{center}
		\caption{\label{fig:colloidal_abs} The calculated absorbance $\alpha(\omega)$ for spherical QDs with various Sn contents ($x$) and diameters ($D$). As the number of states is limited, we truncated the $\hbar \omega$ to: $2.3$~eV, $1.9$~eV, and $1.5$~eV, for the QDs S1, S5, and S9; respectively. }
	\end{figure}
	In this section, we discuss the electronic and spectral properties of spherical Ge$_{1-x}$Sn$_x$ QDs. In this case, the BC2 are imposed. Fig.~\ref{fig:colloidal_en}(a) presents the energies of the two lowest ($e_1$ and $e_2$, neglecting the spin degeneracy) electron states as a function of the dot diameter ($D$) with Sn content fixed at $x = 10$\%. For such a composition, the Ge$_{1-x}$Sn$_x$ alloy is in the regime of the direct band gap. However, the behavior of states in a nanostructure can be very different due to the confinement effect, as discussed in the previous section. 
	In fact, the TB results in Fig.~\ref{fig:colloidal_en}(a) show a very small splitting between the energy levels for the dot diameters up to about $15$~nm, which indicates a dominant role of the $L$-point contributions. The ground state gradually changes its character from about $D$ = 15~nm, where the splitting starts increasing despite the fact that the QD size gets bigger. Furthermore, one can see a change in the energy dependence of the $e_2$ state, which takes place at about $D = 32$~nm. As discussed in the previous section, this behavior is related to the difference in the effective masses that makes the $L$-like states more favorable energetically for a strong confinement regime. The effect is also clearly visible in the composition dependence of the lowest electron energy levels [Fig.~\ref{fig:colloidal_en}(b)]. The splitting between the two lowest states clearly starts increasing when composition reaches the critical value corresponding to direct-indirect transition of the lowest state for a given QD size. Due to the confinement, it takes place at $x \approx 11$\% for $D = 12$~nm and at $x \approx 9$\% at $D = 25$~nm. For the latter, one can also see the transition in the character of the $e_2$ state (for $x \approx 12$\%). \blue{The effect of confinement can be also seen in the band-gap dependencies for GeSn pMOSFETs under tensile strain~\cite{Gupta2014}.}
	
	We compared the tight-binding data with the results obtained from the effective mass approximation (EMA). In the latter model, we assumed the infinite well spherical potential with the energy levels given via the analytic solution~\cite{Fitzpatrick2015}
	\begin{equation*}
		E_{n,l} =  E_c + z^2_{n,l} \frac{2 \hbar^2}{m_{\mathrm{e}} D^2}, 
	\end{equation*}
	where $z_{n,l}$ are zeros of the Bessel functions, $E_c$ is the conduction band edge, and $m_{\mathrm{e}}$ is the electron effective mass. We calculated the energies for $\Gamma$ and $L$ points separately taking $E^{\Gamma}_c = 0.628$~eV and $E^{L}_c = 0.671$~eV, respectively.
	These values were extracted from the results for the Ge$_{1-x}$Sn$_x$ box under the periodic BC1 [Fig.~\ref{fig:box}(a)], where $E^{\Gamma}_c$ can be taken directly from the ground state energy, and $E^{L}_c$ comes from a linear extrapolation of the initial part of the plot.
	The electron effective masses for alloy at $x=10$\% are calculated using the formulas given in Ref.~\cite{Song2019}. For the $L$ point, they result in $m^{\mathrm{L}}_{e,l} = 1.453 m_0$, $m^{\mathrm{L}}_{e,t} = 0.081 m_0$ for the longitudinal and the transversal components, respectively. To obtain an anisotropic parameter, suitable for the spherical model, we averaged $m^{\mathrm{L}}_{e} = 3/( 1/m^{\mathrm{L}}_{e,l} + 2/m^{\mathrm{L}}_{e,t})$. For the $\Gamma$ point, the value is $m^{\mathrm{\Gamma}}_e = 0.0288 m_0$.
	 As shown in Fig.~\ref{fig:colloidal_en}(a), for the $\Gamma$-point regime, the effective mass model significantly overestimates the ground state energy $e_1$ and the splitting between the $e_2$ and $e_1$ energy levels. Similar behavior can be seen in the results of Ref.~\cite{Lee2009} for Ge nanocrystals.  One of the factors may be the deviation from the ideal $\propto k^2$ dependence for the CB near $\Gamma$ in the relevant energy region. Furthermore, the dangling bonds passivation scheme in the TB approach differs significantly from the infinite potential case.
	 On the other hand, for the considered $x=10$\%, we obtained a good agreement between the TB and the EMA models for the energy branches related to the dominant $L$-like contributions. \blue{The general limitations of EMA models are discussed in Ref.~\onlinecite{Mielnik-Pyszczorski2018}.}
	
	We also study the optical properties of colloidal quantum dots. We consider three experimentally-relevant QD compositions and sizes:  Sn content $x=1.5$~\% with the diameter $D = 3.3$~nm (S1); $x=4.2$~\% with $D = 3.9$~nm (S5); and $x=9.1$~\% with $D = 5.0$~nm (S9). These parameters refer to the average values extracted from the samples $1$, $5$, and $9$ from Ref.~\cite{Tallapally2018}. In that experimental work, reflectance spectra were recorded and converted to absorbance within the Kubelka-Munk approach~\cite{Nowak2009}; also the photoluminescence spectra are provided. For the considered samples $1$, $5$ and $9$; the authors obtain (by fitting) the absorption edges: $1.72$~eV, $1.30$~eV, and $0.84$~eV, respectively. Since these values are very different to the corresponding alloy energy gaps (for a given Sn content), the authors suggest an important role of the confinement, as can be expected owing to the relatively small sizes of the QDs. We note that the optical activity of very small (size up to $3$~nm) QDs was studied directly in DFT (HSE functional) approach~\cite{Esteves2016, Hafiz2016}.
	
	The three considered quantum dots are clearly in the indirect band-gap regime, which makes the modeling of their optical properties much more challenging than for well-established InGaAs QD systems. In contrast to the latter, the splittings between the energy levels are very small (creating a quasi-continuum) and the intervalley mixing appears, as discussed above. We calculated the absorbance (from Eq.~\ref{eq:absPropTo}) for a large portion of the spectrum. To this end, we found the 300 lowest cb. and the 200 highest vb. states, and we calculated oscillator strengths ($f_{mn}$) between the pairs of states (but neglecting the intraband transtions inside the VB and the CB). The energy differences between the electron and hole states are shifted down by the values of Coulomb matrix elements, as discussed in Sec.~\ref{sec:abs}.	We arbitrarily take the broadening parameter as $\hbar \gamma = 50$~meV.
	
	The resulting spectra are shown in Fig.~\ref{fig:colloidal_abs}. The absorption edge crucially depends on the QD composition and size, since they determine the band edges and the confinement. The structure of the peaks may to some extent reflect the ladder of discrete states built by the bright $\Gamma$-point-like electron-hole configurations. However, the systematic analysis of this effect is difficult due to the intervalley hybridization, as already discussed. Some signs of the peak-like structure may be also visible in the experimental results of Ref.~\cite{Tallapally2018}, although these come from the samples containing many QDs and are taken at room temperature. Considering the absorption edges, the calculated spectra present a satisfactory agreement with the experimental data. One should keep in mind that experimental data were recorded at room temperature, while the theoretical results correspond to $0$~K. For bulk Ge, the temperature-related shift in the energy gap at the $\Gamma$ point is of the order of 90meV~\cite{Bandstru68:online}.
	
	\section{Conclusions}
	\label{sec:conclusions}
	We found a consistent parameter set for sp$^3$d$^5$s$^*$ tight binding model, for Ge, Sn, and GeSn crystals; and used it to model Ge$_{1-x}$Sn$_x$ alloys and nanostructures. The electron and hole states were calculated in cubic and spherical QDs. We discussed the impact of quantum confinement on the transition from the indirect to the direct band gap regime.  We identified the critical system sizes for such a transition and showed that for small QD systems, this takes place for considerably higher Sn contents, compared to the alloy case. Finally, we calculated absorbance spectra for Ge$_{1-x}$Sn$_x$ spherical quantum dots and obtained a satisfactory agreement with the experimental data.

	\appendix
	\section{Comparison to the literature results}
	
	\begin{figure}[tb!!]
		\begin{center}
			\includegraphics[width=.48\textwidth]{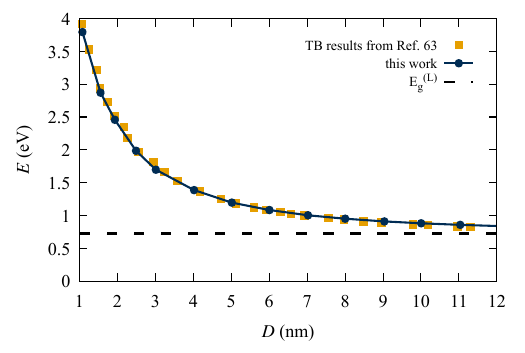}
		\end{center}
		\caption{\label{fig_app:compar} \blue{The single-particle energy gap for Ge spherical quantum dot as a function of diameter. The comparison between the results from Ref.~\onlinecite{Niquet2000} and our paper. The dashed line shows the indirect band gap for Ge bulk crystal.}}
	\end{figure}
	\blue{We compare our results for spherical Ge quantum dot with existing results from the literature. We extracted data points (using WebPlotDigitizer~\cite{WebPlotDigitizer}) related to the sp$^3$d$^5$s$^*$ TB results from Ref.~\onlinecite{Niquet2000} and compared them to our results. Fig.~\ref{fig_app:compar} shows the results for the energy difference between the lowest electron state, and the lowest hole state (the highest valence band state). Following Ref.~\onlinecite{Niquet2000}, the QD diameter is calculated from the number of atoms $N_a$ in the system with the formula $D = 0.351 (N_a)^{1/3}$. Despite different parameter sets used in the calculations, the results are in a very good agreement. We also note that for the Ge quantum dot, Ref.~\onlinecite{Niquet2000} demonstrates that the TB model gives better agreement to the near-infrared PL data, compared to the $\kp$ method.}

        \acknowledgments
	We acknowledge the support from the Polish National Science Centre based on Decision No. 2016/21/B/ST7/01267.
	The calculations for nanostructures have been carried out using
	resources provided by Wroclaw Centre
	for Networking and Supercomputing (\url
	{http://wcss.pl}), Grant No.~203, while the DFT calculations have been performed under Grant No.~45.

	\section*{Conflict of Interest}
	The authors have no conflicts to disclose.

	\section*{Author Contributions}
\textbf{K. Gawarecki}: Conceptualization (equal), Methodology (lead), Software (lead), Formal analysis (lead), Investigation (lead), Writing -- Original Draft (lead), Visualization (lead);
\textbf{J. Ziembicki}: Methodology (supporting), Software (supporting), Formal analysis (supporting), Investigation (supporting), Writing -- Original Draft (supporting);
\textbf{P. Scharoch}: Investigation (supporting), Methodology (supporting), Supervision (lead), Review \& Editing (equal), Funding acquisition (lead);
\textbf{R. Kudrawiec}: Conceptualization (equal), Review \& Editing (equal), Validation (supporting)

	\section*{Data Availability}
	The data that support the findings of this study are available
	from the corresponding author upon reasonable request.

\bibliography{abbr,./library.bib, krzysiek.bib, Ref_DFT.bib}

\end{document}